\documentclass[aoas,MSNbibl,nameyear,seceqn,dvips]{arximspdf}
\usepackage{graphicx}
\usepackage{url,breakurl}
\doi{10.1214/14-AOAS740} 
\volume{8}
\issue{3}
\pubyear{2014}
\firstpage{1561}
\lastpage{1582}

\makeatletter
\newcommand{\rrVert}{\Vert}
\newcommand{\llVert}{\Vert}
\makeatother

\begin{document}
\begin{frontmatter}

\title{Time-warped growth processes, with applications to the modeling of boom--bust cycles in house prices}
\runtitle{Time-warped growth processes}

\begin{aug}
\author[A]{\fnms{Jie} \snm{Peng}\corref{}\thanksref{T1}\ead[label=e1]{jipeng@ucdavis.edu}},
\author[A]{\fnms{Debashis} \snm{Paul}\thanksref{T2}\ead[label=e2]{debpaul@ucdavis.edu}}
\and
\author[A]{\fnms{Hans-Georg} \snm{M\"uller}\thanksref{T3}\ead[label=e3]{hgmueller@ucdavis.edu}}
\runauthor{J. Peng, D. Paul and H.-G. M\"uller}
\affiliation{University of California, Davis}
\address[A]{Department of Statistics\\
University of California, Davis\\
One Shields Avenue\\
Davis, California 95616\\
USA\\
\printead{e1}\\
\phantom{E-mail:\ }\printead*{e2}\\
\phantom{E-mail:\ }\printead*{e3}}
\end{aug}
\thankstext{T1}{Supported in part by National Science Foundation Grant
DMS-10-07583.}
\thankstext{T2}{Supported in part by National Science Foundation Grant
DMS-11-06690.}
\thankstext{T3}{Supported in part by National Science Foundation
Grants DMS-11-04426 and DMS-12-28369.}

\received{\smonth{12} \syear{2013}}
\revised{\smonth{3} \syear{2014}}

%
\begin{abstract}
House price increases have been steady over much of the last 40 years, but
there have been occasional declines, most notably in the recent housing bust
that started around 2007, on the heels of the preceding housing bubble. We
introduce a novel growth model that is motivated by time-warping models in
functional data analysis and includes a nonmonotone time-warping component
that allows the inclusion and description of boom--bust cycles and facilitates
insights into the dynamics of asset bubbles. The underlying idea is to model
longitudinal growth trajectories for house prices and other phenomena, where
temporal setbacks and deflation may be encountered, by decomposing such
trajectories into two components. A first component corresponds to underlying
steady growth driven by inflation that anchors the observed
trajectories on a
simple first order linear differential equation, while a second boom--bust
component is implemented as time warping. Time warping is a commonly
encountered phenomenon and reflects random variation along the time
axis. Our
approach to time warping is more general than previous approaches by admitting
the inclusion of nonmonotone warping functions. The anchoring of the
trajectories on an underlying linear dynamic system also makes the time-warping
component identifiable and enables straightforward estimation
procedures for all
model components. The application to the dynamics of housing prices as
observed for 19 metropolitan areas in the U.S. from December 1998 to July 2013
reveals that
the time setbacks corresponding to nonmonotone time warping
vary substantially across markets and we find indications that they are
related to market-specific growth rates.
\end{abstract}

%
\end{frontmatter}

\section{Introduction}\label{sec1}\label{secintro}

House price increases have been steady over much of the last 40 years, but
there have been occasional declines, most notably in the recent housing bust
that started around 2007. If underlying inflation was acting without any
additional market forces, this would imply a steady rate of increase in house
prices such that log prices are linearly increasing with inflation. However,
asset prices typically do not follow a simple growth model with steady price
increases, but rather are subject to occasional wild oscillations, as
exemplified by the recent U.S. housing boom and bust cycle that led to
a major
worldwide financial crisis. Such asset price swings have been
attributed to
irrational and herd behavior of investors by Shiller, who identified and
described these forces [\citeauthor{shill05} (\citeyear{shill05,shill08,shill13})]
which render housing
markets inefficient [\citet{case89}]. Shiller received a Nobel
prize in economics
in 2013 for this work.

It is therefore of substantial interest to determine asset price
dynamics in
time windows around asset bubbles, which have been a historically recurring
phenomenon, in order to understand which features capture and describe the
extent and dynamics of bubbles and busts, a topic that has found recurring
interest [\citet{bondt1985does,case03}, \citeauthor{yan2012diagnosis} (\citeyear{yan2012diagnosis,yan2012inferring})]. The
idea of an underlying smooth and stable growth trajectory in asset
prices motivates a
model that includes an underlying first order linear differential
equation with
a market-specific growth rate, which by itself reflects steady exponential
growth. This component then needs to be complemented by a boom--bust component,
for which we employ nonparametric time warping. The idea is that a
price bust leads
to a setback in time, while a boom leads to an acceleration in the way time
moves forward, as prices move faster into the future than the actual
flow of
chronological time.

Colloquially, a setback in time when a bust occurs is
reflected in statements that house or other asset prices are currently at
levels that correspond to those of a past calendar year. Thus, we model
longitudinal growth trajectories for house prices and other phenomena where
temporal setbacks may be encountered, such as other asset prices or weight
increases in growing organisms, by decomposing the observed growth into an
underlying steady growth component that anchors the observed
trajectories in a
simple linear system, and time warping to reflect price swings. The
application of our approach for asset price modeling to housing prices, as
observed for 19 metropolitan areas in the U.S. from December 1998 to
July 2013,
reveals that the amounts of time setbacks between the markets vary
substantially and are related to underlying growth features.

We view the price curves observed for various markets as a sample of functional
data. Such data have become commonplace in many fields, including chemometrics,
econometrics, etc. [\citet{ferraty2006nonparametric,rams022}].
As features not only vary in terms of amplitude, but
also in time of occurrence, time
variability across curves is a common observation. For example, when
considering biological growth, humans achieve maximum growth velocity
at a
subject-specific age, as subjects progress to and through puberty at different
ages. Time warping (also known as registration or alignment) aims to address
this variability by transforming the time domain of each function, normally
under the constraint of monotonicity. Curve alignment is also motivated
by the
belief that for many systems, it is a (subject specific) intrinsic
time, rather
than clock time, that governs the underlying dynamics. In such
situations, the
explicit inclusion of time warping in functional data models leads to reduced
variability and better interpretation. In Functional Data Analysis the
presence of time warping is often considered a nuisance. In contrast,
for the
case of asset prices, time warping corresponds to a component reflecting
price swings and bubbles and therefore is a key part of asset price
modeling. In usual functional data settings, the presence of time warping
routinely leads to identifiability problems, as time and amplitude variation
are generally not separable in functional data settings, where one aims at
modeling a sample of random functions [\citet
{kneip1992statistical,wang99,mull044}].

A popular approach for curve alignment is the landmark method where one
defines a set of landmarks, which then are time transformed so as to
occur at
the same transformed time points across curves [\citet{sakoe1978dynamic,kneip1988convergence}]. More recent work on curve alignment and registration
includes \citet{james2007curve} (curve alignment by moments),
\citet{knei08}
(alignment and functional principal components), \citet{tele08} (Bayesian
approaches to time warping), \citet{mull086} (inferring global
registration
from pairwise warping), \citet{sangalli2010} ($k$-means alignment
for curve
clustering) and \citet{srivastava2011registration} (registration
of functional
data with the Fisher--Rao metric).

\begin{figure}

\includegraphics{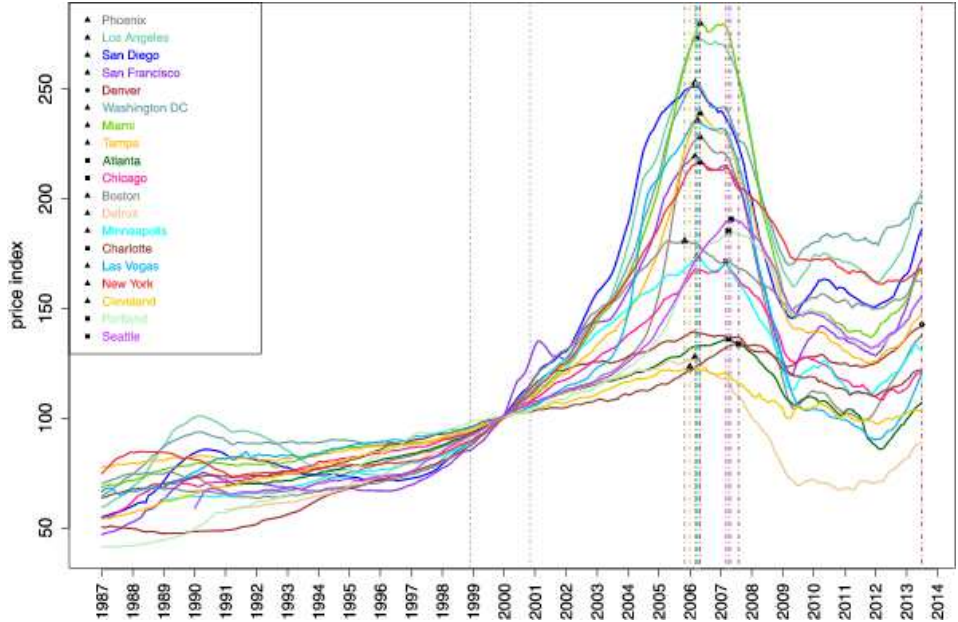}

\caption{Housing price indices for 19 metropolitan areas in the U.S.
from January
1987 to July 2013. These indices are three-month moving averages and are
normalized to have a value of 100 in January 2000 (for more details, see
Section~\protect\ref{sec3}).} \label{figurehouseindex}
\end{figure}

Our proposed approach to time warping is more general than previous approaches
in two key respects: first, it allows for inclusion of nonmonotone
\mbox{time-}warping functions, an essential feature for modeling the busts
that occur
in boom--bust cycles, since these correspond to a setback in time,
with prices recurring to those of a past period. Second, it overcomes
the usual
identifiability problems of the time-warping component by anchoring the
trajectories to an underlying linear dynamic system. This anchoring
makes it
possible to introduce straightforward estimation procedures for all model
components. While the central interest of this paper is the analysis of price
oscillations in housing markets, our model extends beyond housing
prices to
other systems for which an underlying growth rate may be assumed, such
as long-term
behavior of the stock market or similar markets, for which long-term
appreciation rates are meaningful [\citet{bondt1985does}].

As an illustration that growth may substantially deviate from an exponential
trajectory (e.g., during an economic bubble) or growth may even become negative
(e.g., during the burst of a bubble), Figure~\ref{figurehouseindex} shows
seasonally adjusted S\&P/Case--Shiller Home Price Indices for $19$ metropolitan
areas in the United States from 1987 to 2013. As can be seen from this figure,
the housing price trajectories do not correspond to exponential curves,
especially
after year 2000. Indeed, many areas enjoyed higher growth rates from
2000 to
2006, as compared to the 1990s. However, this accelerated growth was
followed by a
bust, a sharp decline of housing prices after 2006/2007.

\begin{figure}

\includegraphics{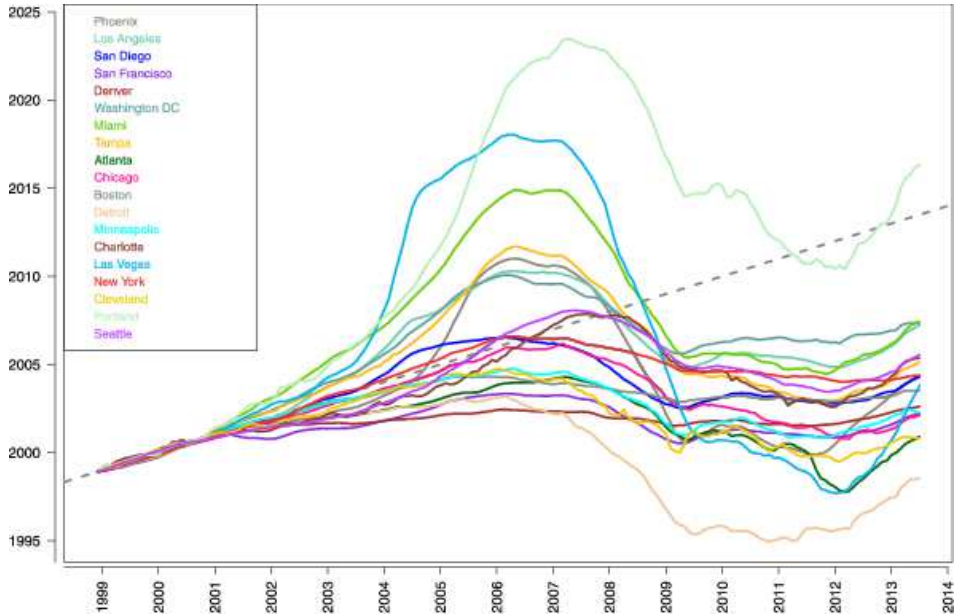}

\caption{Warping functions for 19 metropolitan areas in the U.S. from December
1998 to July 2013. The broken grey line represents the identity function
$h(t)=t$.}\vspace*{-3pt} \label{figurerawwarptogether}
\end{figure}

\begin{figure}

\includegraphics{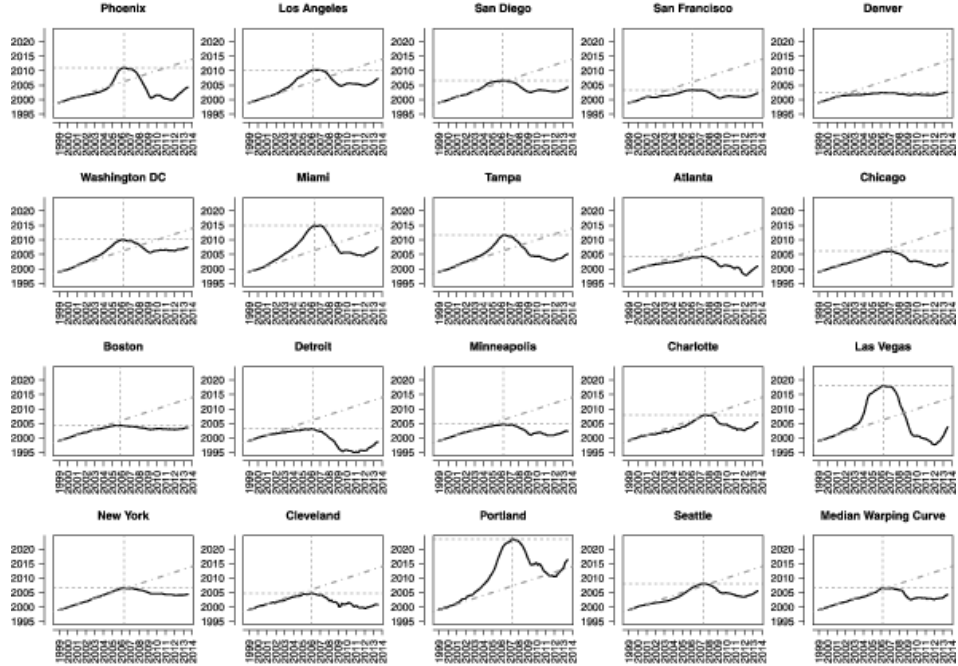}

\caption{Time-warping functions for 19 metropolitan areas in the U.S. from
December 1998 to July 2013 (drawn in individual panels), mapping chronological
time (on the abscissa) to market time (ordinate). Dashed grey lines represent
the identity function $h(t)=t$. Vertical grey lines indicate the timing
of peak
price, while horizontal grey lines indicate the level of peak price.}
\label{figurewarpraw}
\end{figure}

Even though the general trends are somewhat similar across the metropolitan
areas, the timing of the peak price, periods of fastest growth and
rates of
decline after the peak, etc., turn out to be quite variable. Coupled
with the
underlying linear dynamics, these variations can be accounted for by
area-specific warping effects. Figures~\ref{figurerawwarptogether} and
\ref{figurewarpraw} show the warping functions for the time period
1998 to
2013 derived from the model proposed in Section~\ref{sec2}. It can be seen that housing
prices first warped forward in time. After the housing market collapsed around
2006/2007, the housing price warped backward in time and then started
to move
forward again around 2012. But for all markets except for Portland,
Oregon, the
warped time remains substantially below the calendar time until the present.
Interestingly, these time setbacks vary quite a bit between different
housing markets. The time setback of housing prices is consistent with the
notion of an economic time reset, which has been featured in many
discussions since
the onset of the recent recession.

In addition to mapping asset price development on a market-specific timeline,
our analysis and methodology identifies boom and bust components in the
time-warping functions and positions the different housing markets on a
boom--bust
plane that indicates to what extent specific markets reflect boom or
bust to a
larger or lesser extent. We also explore relationships of booms and
busts with underlying
steady growth rates. The basic model is introduced in the next section,
and a study of boom--bust cycles follows in Section~\ref{sec3}.
Our analysis is supported by simulation results that are reported in
Section~\ref{sec4} and theoretical considerations that are in the \hyperref[app]{Appendix}.
The paper ends with a discussion in Section~\ref{sec5}.

\section{Time-warped growth model}\label{sec2}\label{secmethod}

Our approach is to model the observed continuous time asset price trajectories
through an underlying linear dynamical system, coupled with a random
time-warping component. The $i$th trajectory $X_i(\cdot)$ that
corresponds to the
housing price curve of a city, as quantified by the Case--Shiller house price
index, GDP per capita curve of a country or stock market index, is
assumed to
result from time warping due to ``irrational'' market forces that act in
conjunction with an underlying ``smooth growth'' process $Z_i(\cdot)$ that
corresponds to an underlying market-specific rate of appreciation.
Scaling the
observation time to $[0,1]$,
%
\begin{equation}
\label{eqtimewarp} X_i(t) = Z_i\bigl(h_i(t)
\bigr),\qquad t \in[0,1], i=1,\ldots, n,
\end{equation}
where $h_i(t)$ is a market-specific warping function, a key component
of our
model. Furthermore, the underlying ``smooth'' growth process $Z_i(\cdot
)$ that
reflects market-specific long-term growth is assumed to follow a first order
linear ordinary differential equation, that is,
%
\begin{equation}
\label{eqlineardyn} Z_i'(t) = \alpha_i
Z_i(t),\qquad t \in[0,1],  i=1,\ldots, n.
\end{equation}
Here $\alpha_i>0$ is a market-specific random effect that captures the
intrinsic growth rate of the $i$th market.


From equations (\ref{eqtimewarp}) and (\ref{eqlineardyn}), we have
%
\begin{equation}
\label{eqXdev} X_i'(t)= Z_i'
\bigl(h_i(t)\bigr) h_i'(t) =
\alpha_i Z_i\bigl(h_i(t)\bigr)
h_i'(t) = \alpha _i X_i(t)
h_i'(t),
\end{equation}
whence,
\[
\alpha_i h_i'(t) = \frac{X_i'(t)}{X_i(t)} =
\frac{d}{d t} \log\bigl(X_i(t)\bigr).
\]
If we assume $h_i(0)=0$ ($i=1,\ldots, n$), then
%
\begin{equation}
\label{eqwarpfunctions} h_i(t)= \frac{1}{\alpha_i} \log\frac{X_i(t)}{X_i(0)},\qquad
t \in [0,1],  i=1,\ldots, n.
\end{equation}

As market prices $X_i(\cdot)$ generally tend to increase in the long
term but
are nonmonotonic in the shorter term (as demonstrated by the housing price
trajectories in Figure~\ref{figurehouseindex}), the warping functions
$h_i(\cdot)$ will be nonmonotonic.
Moreover, if for some $0\leq t \leq1$,
$X_i(t)<X_i(0)$, then $h_i(t)$ will be negative. We interpret decreasing
warping functions as time moving backward, while increasing warping functions
would signal time moving forward. When a warping function is negative or
greater than $1$, it is interpreted as the system having warped back to the
past or having leapt forward to the future beyond the time interval
where the
sample curves are being observed, respectively.

From equation (\ref{eqwarpfunctions}), the warping function
$h_i(\cdot)$ is
determined not only by the observed process $X_i(\cdot)$, but also by the
intrinsic growth rate $\alpha_i$. The model specified by equations
(\ref{eqtimewarp}) and (\ref{eqlineardyn}) is therefore identifiable up to
$\alpha_ih_i(t)$. A plausible assumption that we will make to render the
model fully identifiable is that the observed trajectory $X_i(\cdot)$ follows
a linear dynamical system on a small time interval, say, $[0, t_0]$, where
growth is smooth, without price oscillations, so that there is no disturbance
from the random warping and the warping function on this interval is
$h_i(t)=t,  t \in[0,t_0]$. Then we may model $X_i(\cdot)$ as
%
\begin{equation}
\label{eqidentify} X'_i(t)=\alpha_i
X_i(t),\qquad t \in[0, t_0].
\end{equation}
For example, from Figure~\ref{figurehouseindex}, we find that the housing
prices followed nearly exponential growth paths in the late 1990s/early 2000,
and the growth rates $\alpha_i$ can then be recovered from this time interval.

Specifically, from (\ref{eqXdev}), it is easy to see
\[
\alpha_ih_i'(t)=\alpha_i,\qquad t \in[0, t_0],
\]
so that $h_i(0)=0$ implies
%
\begin{equation}
\label{eqwarpexp} h_i(t) = t,\qquad t \in [0, t_0].
\end{equation}
Since on $[0, t_0]$, $X_i(t)=C\exp(\alpha_i t)$, $\alpha_i$ is
determined by
\[
\alpha_i =\frac{1}{t}\log\frac{X_i(t)}{X_i(0)},\qquad t \in(0,
t_0]
\]
or, equivalently,
%
\begin{equation}
\label{eqalpha} \log X_i(t) = \log X_i(0) +
\alpha_i t,\qquad t \in[0, t_0].
\end{equation}
Under the model specified by (\ref{eqtimewarp}), (\ref
{eqlineardyn}) and
(\ref{eqidentify}), $\alpha_i$ is interpreted as the rate of growth
on the
time period $[0, t_0]$ and the warping function $h_i(\cdot)$ captures the
(possible) deviation of the system on the time period $[t_0,1]$ from
the linear
dynamics on the time period $[0, t_0]$.

From now on, we work within the framework of this model, as we are primarily
interested in identifying the patterns of the most recent housing
market cycle.
The time period $[0,t_0]$ is chosen as a two-year period in the late
1990s/early 2000s; more discussion on this choice follows below. Our
goal is
to study the patterns of house price oscillations during the past 10 years,
contrasting it to a smooth price growth prior to this period.

In practical applications, data may not always follow the model
exactly, so
that equation (\ref{eqalpha}) only holds approximately. Then one may
obtain an
underlying smooth growth rate $\alpha_i$ by minimizing a sum of
squares type
criterion,
%
\begin{equation}
\label{eqalphaest} \alpha_i=\operatorname{arg}\min_{\alpha_i>0} \int
_{ [0, t_0]} \bigl(\log X_{i}(t)-\log X_i(0)
-a_i t \bigr)^2 \,dt,
\end{equation}
where integrals are approximated in practice by appropriate Riemannian
sums. We
adopted this approach for the analysis of the housing price index data
in the
next section, where the interval $[0, t_0]$ is chosen so as to maximize the
coefficient of determination when fitting model (\ref{eqalpha}) to
the data.
Another option to determine the rate $\alpha_i$ would be to use external
information, such as a historic market rate of price appreciation,
which may be
tied in some way to the development of rents or inflation under the
belief that
the ``rational'' increase in house prices would match this underlying growth
rate in the future [\citet{shill05}].

Note that the $\alpha_i$ serve as nuisance
parameters and are not of interest in themselves, in contrast to the warping
functions $h_i$, which capture the departures from the ``rational'' growth
rate in the presence of price swings.
Once growth rates $\alpha_i$ are obtained, by (\ref
{eqwarpfunctions}) one
arrives at the time-warping functions
%
\begin{equation}
\label{eqwarpraw} h_i(t)=\frac{1}{\alpha_i}\bigl(\log
X_i(t)- \log X_i(0)\bigr),\qquad i=1,\ldots, n.
\end{equation}

We then determine the main modes of variation of the warping functions through
functional principal component analysis (FPCA). This method has evolved
into a
powerful tool of functional data analysis to summarize functional data
and to
capture their variation [\citet{cast86,rice91,RamSil,yao2005,peng2009geometric}].
Starting with a sample of time-warping functions, this method determines
eigenfunctions $\phi_k,   k \ge1$, of the auto-covariance operator
of the
underlying warping process, which forms an orthonormal basis of
function space,
as well as the associated eigenvalues $\lambda_k,  k \ge1$, which
provide an
indication of the fraction of variance that is explained by a particular
eigenfunction. Additionally, one obtains the functional principal components
(FPCs) $\xi_{ik}$, which are random scores that correspond to the expansion
coefficients in the eigenbasis, that is, the basis formed by the eigenfunctions.
The FPCs are obtained by projecting centered time-warping functions on the
$k$th eigenfunction, $k \ge1$. FPCA provides a parsimonious representation
of the data and achieves efficient dimension reduction. Further details about
the methodology and consistency results can be found in the \hyperref[app]{Appendix}.

A tool to visualize FPCA that we employ for the housing price data is
the modes
of variation plot [\citet{jone92}]. In these plots the direction
in function space,
into which a given eigenfunction points, is visualized by $\mu(t)+
\gamma
\sqrt{\lambda}_k \phi_k(t),  t \in[0,1]$, where one varies $\gamma
$ over a
range of values, typically $\gamma\in[-2,2]$. This plot provides a
visual indication of the movement
from the mean function $\mu$ toward the positive and negative
direction of the
$k$th eigenfunction $\phi_k$.

\section{Boom and bust in U.S. housing markets}\label{sec3}\label{sechousing}

Here, we apply the method described in Section~\ref{sec2} to study the housing price
trends in 19 U.S. metropolitan areas. The data were downloaded from 
\url{http://www.spindices.com/index-family/real-estate/sp-case-shiller}.
The original data consist of seasonally adjusted Case--Shiller Home Price
Indices [\citet{case87}] for each month for 20 metropolitan
areas and two composite
indices from January 1987 to July 2013. These indices are three-month moving
averages and are normalized to have a value of $100$ in January 2000.
For more details about how the housing indices are calculated, see ``S\&P/Case--Shiller Home Price Indices Methodology'' published by Standard
\& Poor's. Since for
Dallas, Texas, housing price indices are only available after 2000, we
drop it
from our analysis and thus focus on the remaining $19$ metropolitan areas.

The housing price index trajectories for these 19 areas are depicted in Figure~\ref{figurehouseindex}. For some areas, there is a clear bubble in the
housing market after year 2000, in the sense that housing prices increased
super-exponentially. This phenomenon is clearly illustrated by the warping
functions depicted in Figure~\ref{figurewarpraw}. Also, the timing
of the
``peak price'' (indicated by the dashed colored lines in Figure~\ref{figurehouseindex}) is mainly clustered into two groups: a
larger group
of $13$ areas (triangles) where peak price occurred around late 2005
and early
2006, and a smaller group of $5$ areas (squares) where peak price occurred
around the first two quarters of 2007. The only exception is Denver, Colorado
(diamond), where the summer 2013 indices are slightly higher than the
pre-recession peak that occurred in Febuary/March, 2006.

\begin{figure}

\includegraphics{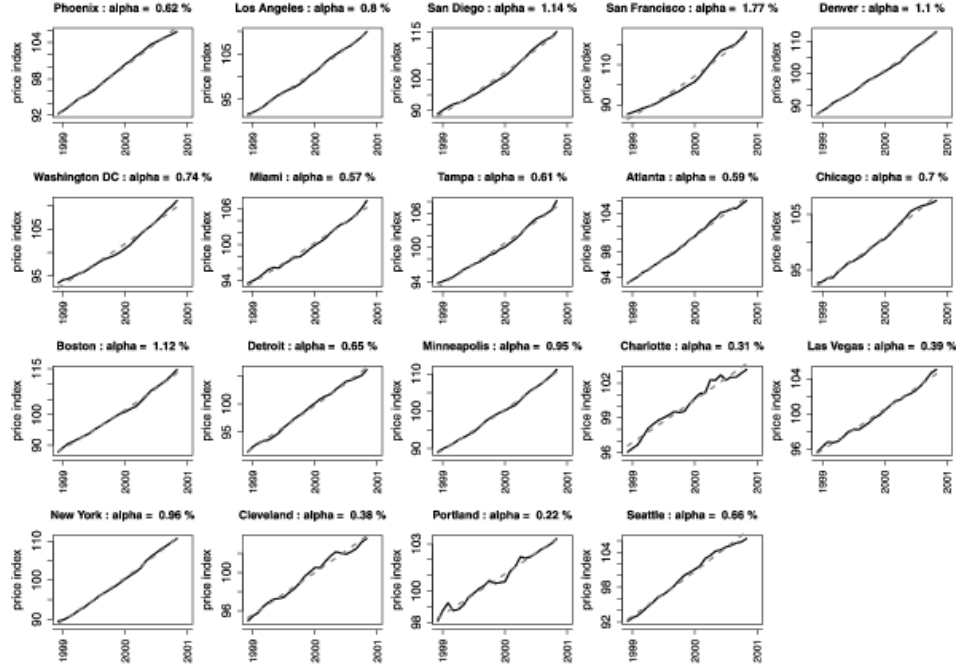}

\caption{Exponential fit (broken grey lines) of housing price indices
(solid lines) between December 1998 and November 2000, where the
$\alpha_i$ are expressed as
$\%$ per month.} \label{figureexpfit}
\end{figure}

We then fitted exponential curves to all 2-year, 3-year, 5-year time intervals
within 1991 to 2013 (we started from 1991 since there is no missing
data after that) and found that for the 2-year interval from December 1998
to November 2000 (the period between the two dotted grey lines on Figure~\ref{figurehouseindex}), the exponential fits result in the largest overall
coefficient of determination $R^2$. As can be seen from Figure~\ref{figureexpfit}, the exponential curve fits the housing price trajectory
between December 1998 and November 2000 very well. More specifically,
on this
interval, among the fits for these $19$ markets, the smallest $R^2$ is $0.967$
(Charlotte) and the largest $R^2$ is $0.998$ (Denver). Since our major
interest is in investigating the boom--bust cycles of the housing market
in recent years, we choose December $1998$ as our starting time.
Specifically, we
treat December 1998 as time $0$, and July 2013 as time $1$ in the normalized
time scale. We also elect $t_0$ to correspond to November 2000 and use $[0,
t_0]$ to estimate the growth rates $\alpha_i$ by equation
(\ref{eqalphaest}). The average growth rate of these $19$
metropolitan areas
is 0.75\% per month, with a standard deviation of $0.37\%$ per month.

It can be seen that some markets, including San Diego, San Francisco, Denver,
Boston, Minneapolis and New York, already had a fast growth rate at the
end of
the last century (with an $\alpha_i$ around or above $1\%$ per month). Figure~\ref{figurewarpraw} shows the warping function for each area defined by
equation (\ref{eqwarpraw}), as well as the median warping curve across
these 19 markets (in the last panel). One finds that the markets already
growing fast at the end of the last century mostly retained a similar growth
rate after 2000 until the housing market collapsed around 2006/2007. Some
markets, including Washington DC, Miami, Tampa, Las Vegas and Portland, went
through a much faster price growth period after 2000. The housing price trend
of Detroit is quite unique. This market had moderate growth at the end of
last century ($\alpha_i=0.65\%$), followed by declines toward 2000 and
a sharp
drop after 2006, which places it in a separate category from the rest.

\begin{figure}

\includegraphics{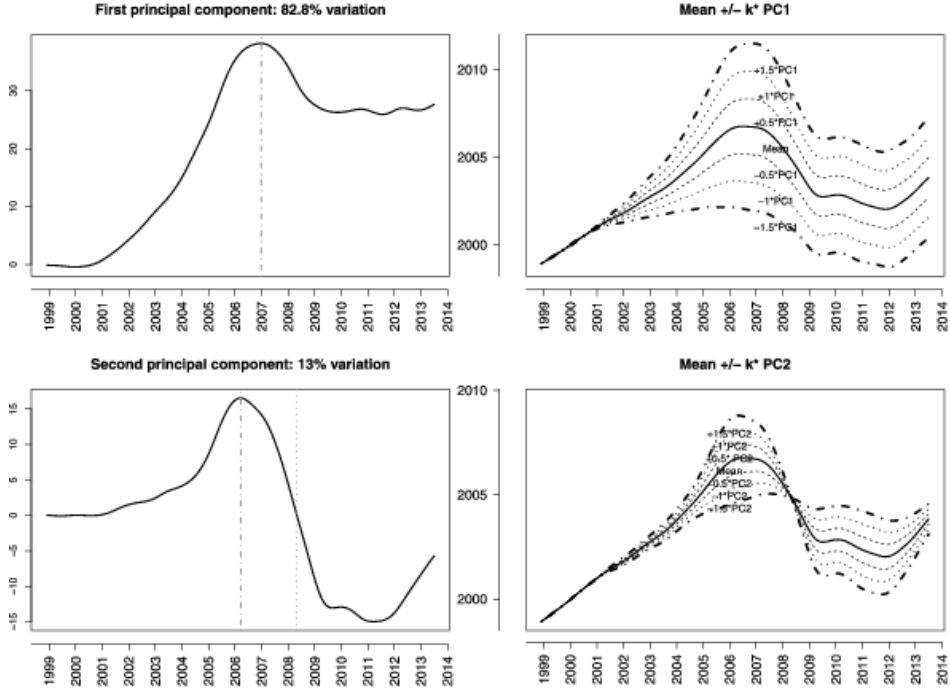}

\caption{First two eigenfunctions (multiplied by the square-root of
their respective eigenvalues) of the time-warping process.}\label{figurewarppc12}
\end{figure}

We then apply FPCA for the warping functions, as described in Section~\ref{secmethod}. The first two principal components explain $ \sim
96\%$ of
the total variation in the warping functions. However, Las Vegas and Portland
appear to be outliers. In order to avoid the influence of these
outliers on our
analysis, we dropped Las Vegas and Portland and redid the FPCA for the
remaining $17$ markets. Again, the first two principal components explain
nearly $96\%$ of the total variation (1st PC: $82.8\%$, 2nd PC $13\%$). Figure~\ref{figurewarppc12} shows the first two eigenfunctions (multiplied
by the
square-root of their respective eigenvalues), which display some remarkable
features. First, the first eigenfunction has a shape that is similar to that
of the mean function. This means a large fraction of the variance is explained
by what amounts roughly to the degree at which the average boom cycle is
expressed.
Second, the first eigenfunction primarily reflects the rise of the house
price to a high peak (around December, 2006) in the boom cycle and the
subsequent fall, but to a level that is still substantially higher than the
2000 price level. Thus, this component primarily reflects the boom part
of the
cycle and is referred to as the ``boom'' component. Third, the second
eigenfunction features an earlier (around March, 2006), smaller but steeper
peak, and then a deep fall to a very low level. Thus, this component primarily
reflects the bust part of the cycle and we call it the ``bust''
component. It
is also interesting to note that the second eigenfunction went from positive
to negative in early $2008$, which coincides with the onset of the economic
recession.

\begin{figure}

\includegraphics{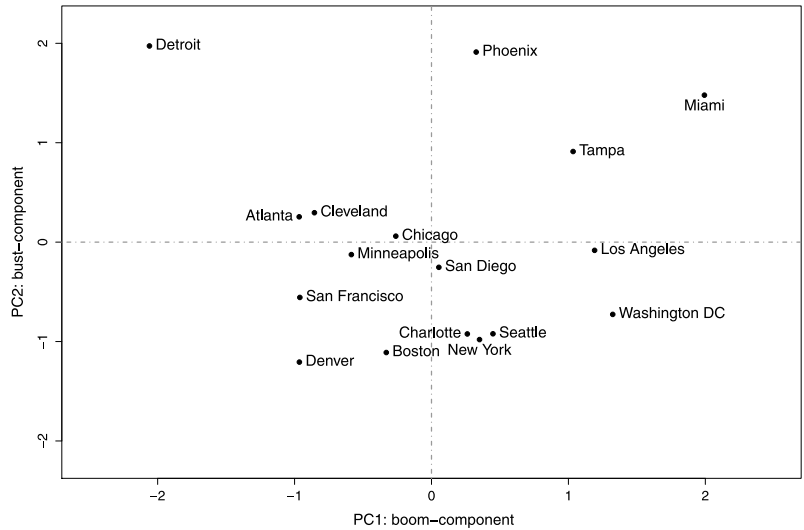}

\caption{First two principal component scores of $17$ metropolitan
areas in the U.S.}
\label{figurewarppc12scores}
\end{figure}

Although the estimation of the growth rates $\alpha_i$ may change when the
fitting interval, on which exponential growth is assumed, varies, we
found that the
FPCA results do not depend much on the choice of the fitting interval.
When we
applied the fitting procedure for the $\alpha_i$ on a different two-year
interval, namely, January 1998 to December 1999, the
resulting eigenfunctions have almost identical shapes: the two leading
principal components explain the majority of the variation in the data,
and the
first component represents a boom component and the second component
corresponds to a bust component. For more details, see the supplementary
material [\citet{supp}].

After extracting the first and second functional principal components (which
are the scalar random variables that are multiplied with the
eigenfunctions in
the representation of the time-warping functions), we plot the ``bust''
component against the ``boom'' component in Figure~\ref{figurewarppc12scores}. The markets falling in the right lower quadrant
(e.g., Washington, DC) experienced a boom but relatively little bust, while
those in the right upper quadrant (e.g., Miami) experienced both boom
and bust
in similar measure. The areas in the left upper quadrant have been
subject to a
bust with little boom (e.g., Detroit), while those in the lower left quadrant
had little boom or bust (e.g., Denver). Even though Las Vegas and Portland
have not been used to derive these components, their projected first
two PC
scores are $1.65$ and $5.57$ for Las Vegas and $4.82$ and $-$0.39 for
Portland. This indicates that in the boom--bust coordinates that emerged from
the FPCA, Las Vegas experienced a big housing boom, followed by an enormous
bust, while Portland enjoyed a huge boom with little bust.

\begin{figure}

\includegraphics{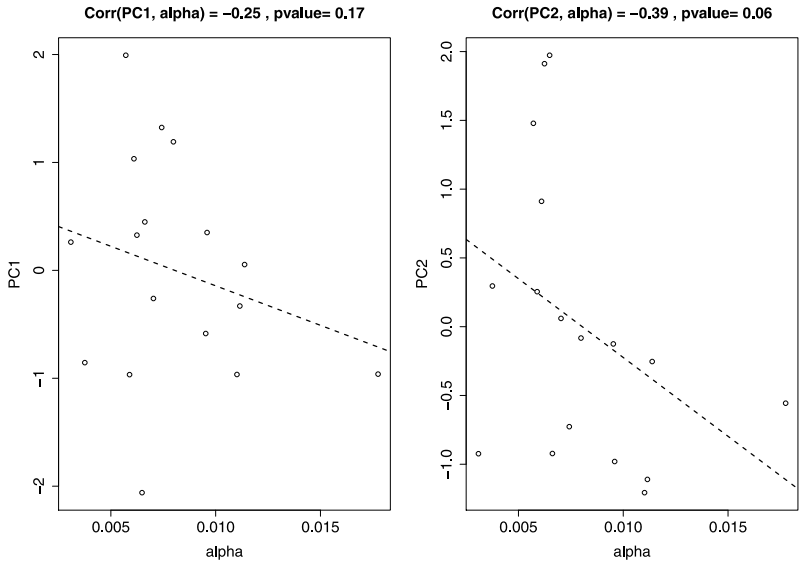}

\caption{Principal component scores versus growth rate $\alpha_i$.}
\label{figurepcvsalpha}
\end{figure}

Also of interest are relations between the time-warping functions, as
characterized by their first two principal component scores, which
serve as
random effects in the eigen-expansion (see the \hyperref[app]{Appendix} for more
details) and the
underlying growth rate $\alpha_i$, which may be viewed as a market-specific
characteristic. The corresponding scatterplots of first (resp.,
second) principal component score versus $\alpha_i$ with the
superimposed least
squares line in Figure~\ref{figurepcvsalpha} demonstrate that
regions with
lower baseline growth rates experienced stronger boom--bust cycles. The
relationship seems particularly striking for the bust
component.

\section{Simulation results}\label{sec4}\label{secsimulation}

In this section we report a simulation experiment that was conducted to
investigate the practical performance of the proposed procedure. This
simulation was designed to mimic the housing index data, and $n$ warping
functions were generated as
\[
h_{i}(t)=\mu(t)+\sum_{k=1}^K
\sqrt{\lambda_k} \xi_{ik} \phi _k(t),\qquad  t
\in[T_0,T_1], i=1,\ldots, n.
\]
The sample size was chosen as $n=20$, the mean function $\mu(\cdot)$
was chosen
as the estimated mean warping function of the housing data (solid curve
in the
upper-right panel of Figure~\ref{figurewarppc12}), $K=10$ and
$\{\phi_k(\cdot)\}_{k=1}^{10}$ were chosen as the first $10$ eigenfunctions
estimated from the housing data (upper-left and lower-left panels of \mbox{Figure}~\ref{figurewarppc12} show the first two eigenfunctions),
furthermore, the
eigenvalues $\{\lambda_k\}_{k=1}^{10}$ were chosen to correspond to
those in
the housing price analysis, and the functional principal component scores
$\xi_{ik}$ were chosen as i.i.d. from $N(0,1)$.
The time interval is set as $T_0=144$ and $T_1=319$, reflecting the monthly
indices from December 1998 to July 2013, respectively (January 1987 is month
$1$).

We then generated the price trajectories according to equation (\ref
{eqwarpfunctions}), namely,
\[
X_i(t)=X_i(T_0)\exp\bigl\{
\alpha_i \bigl(h_i(t)-h_i(T_0)
\bigr)\bigr\},\qquad t \in[T_0, T_1],  i=1,\ldots, n,
\]
where the initial values $X_i(T_0)$ were chosen as i.i.d. from $\operatorname{Uniform}(85, 100)$, while the growth rates $\alpha_i$ were randomly generated
from $\operatorname{Uniform}(0.003,\break  0.018)$, where only samples that
satisfied the boundedness condition\break $\max_{t \in[T_0, T_1]} X_i(t) <
300$ were retained. The
interval (85,100) is set according to the range that was observed for the
housing indices in December 1998, while the interval $(0.003, 0.018)$ matches
the range of the estimated $\alpha_i$ from the housing data. The
trajectories are capped at $300$ since the maximum index in the housing data
is $280$.

\begin{figure}[t]

\includegraphics{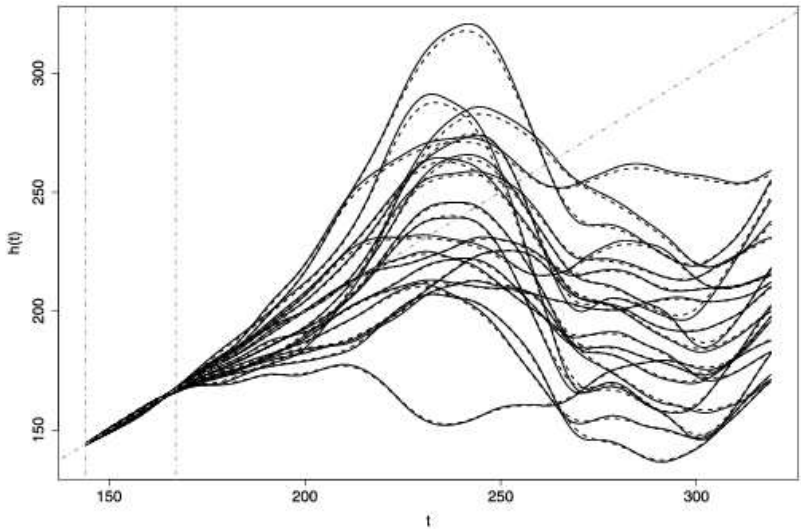}

\caption{Simulated warping functions: true---solid lines, estimated---broken lines. The dashed grey line represents
the identity function $h(t)=t$. Vertical grey lines indicate the end
points of the fitting region for the growth rates used in the real data,
namely, December 1998 (month 144) and November 2000 (month 167).}\vspace*{-6pt}\label{figurewarpsimu}
\end{figure}
%
\begin{figure}[b]

\includegraphics{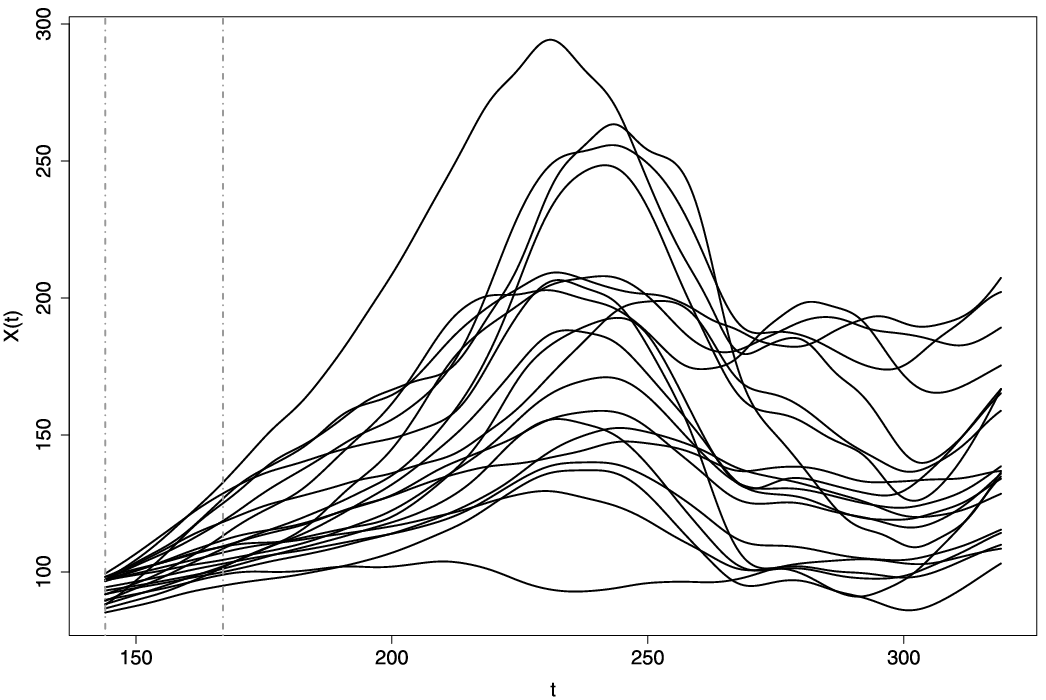}

\caption{Simulated price trajectories. Vertical grey lines indicate
the end
points of the fitting region for the growth rates used in the real data,
namely, December 1998 (month 144) and November 2000 (month 167).}\vspace*{-6pt}\label{figurepricesimu}
\end{figure}

Figure~\ref{figurewarpsimu} shows $20$ simulated warping functions (solid
lines) and Figure~\ref{figurepricesimu} shows the corresponding
simulated price
trajectories. The two vertical grey lines in these figures indicate the end
points of the fitting region for the growth rates used in the real data,
namely, December 1998 (month $144$) and November 2000 (month $167$).

We then applied the same procedures used for analyzing the housing
price index
data. First, we fitted exponential curves on all 24-month, 36-month and
60-month time intervals and calculated the coefficient of determination $R^2$
for each market on each of these intervals. We picked the interval
corresponding to the largest averaged $R^2$ (across the $20$ subjects).
For the
data in Figure~\ref{figurepricesimu}, the best interval coincides
with the
fitting region used in the real data. The corresponding averaged $R^2$ is
$0.997$ and the estimated $\alpha_i$ have an averaged relative squared error
(ASE) $ \frac{1}{n} \sum_{i=1}^n \frac{(\hat{\alpha}_i-\alpha
_i)^2}{\alpha_i^2}
$ of $9.51 \times10^{-5}$. The warping functions were estimated as described
in Section~\ref{sec2} [equation (\ref{eqwarpraw})], and these estimates are displayed
as broken lines in Figure~\ref{figurewarpsimu}. FPCA is then applied
to the
estimated warping functions, and one ends up with an estimate of the
eigencomponents, which are the eigenfunctions and the functional principal
components; see the \hyperref[app]{Appendix} for a more detailed description.

We repeated the above process (including data generation and fitting)
$100$ times.
Across the $100$ replicates, the ASE for estimating the $\alpha_i$ had a
mean of $0.011$ and a standard deviation of $0.041$. The left end point
of the $\alpha$ fitting region was found to have a mean of $146.31$
and a
standard deviation of $7.58$, while the right end point had mean
$169.43$ with
standard deviation $7.93$. The mean relative-integrated-squared-error
for the
warping function estimation was $0.032$. 

\begin{figure}[b]

\includegraphics{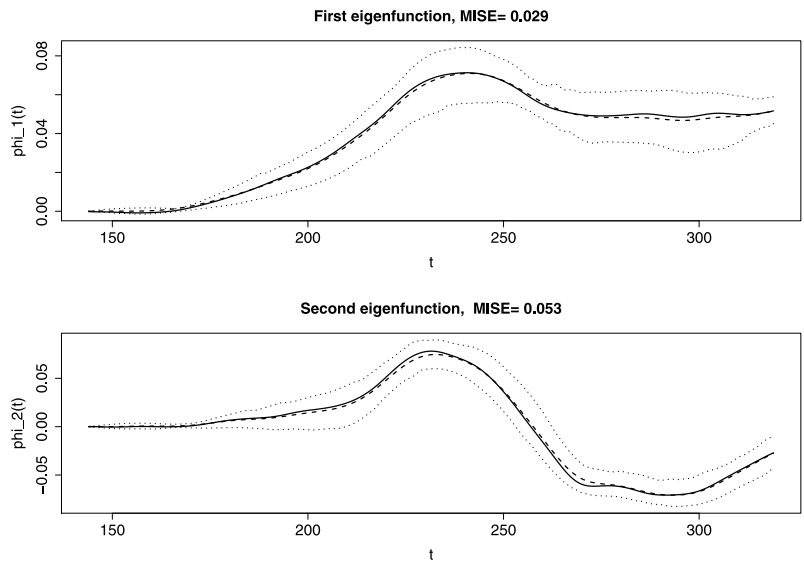}

\caption{First two simulated eigenfunctions: true---solid line,
pointwise mean---broken line, pointwise $95\%$ and $5\%$ bands---dotted lines.} \label{figureeigenestsimu}
\end{figure}

Figure~\ref{figureeigenestsimu} shows the pointwise mean estimated
eigenfunctions (broken lines) and the $95\%$ and $5\%$ pointwise bands
(dotted lines). The mean-integrated-squared-error (MISE), namely, the
average of $\|\hat{\phi} -\phi\|^2$ across the $100$ replicates, is
$0.029$ for
the first eigenfunction and $0.053$ for the second eigenfunction. The mean
relative-squared-error of the first two eigenvalues are $0.825$ and $0.135$,
respectively. The percentage of variation explained by the first two estimated
eigenfunctions ranges from $85.5\%$ to $98.8\%$ with a mean $96\%$.

These results show that the proposed procedure is able to select the right
fitting region for the growth rate $\alpha$ and that it is also
effective in
estimating both $\alpha$ and the warping function $h(\cdot)$. Most
importantly,
as can be seen from Figure~\ref{figureeigenestsimu}, there is
little bias in
the estimation of the leading eigenfunctions. It is particularly noteworthy
that the estimators are able to preserve the shape of the
eigenfunctions (as
reflected by the two bands).

\section{Discussion}\label{sec5}

In this paper we introduce a new model for asset price modeling that is
motivated by and illustrated with the recent price swings observed in
the U.S.
real estate market. Introducing an interface between an underlying dynamic
``steady growth'' model and time warping leads to interesting results and
insights about asset price dynamics. Our approach aims to model asset
prices in
terms of a smooth component that corresponds to a market-specific
steady rate
of appreciation and an oscillation component that reflects ``irrational''
swings in prices and their extreme manifestation in the form of boom--bust
cycles, as recently experienced across the U.S. housing markets. Our analysis
and modeling is not limited to real estate price dynamics and will be of
interest also for the modeling of other asset prices that involve price swings
and boom--bust cycles; this includes equity and art markets.

Key findings from our analysis are that for the recently burst bubble
in U.S.
house prices across several metropolitan area markets, one can clearly
distinguish a boom component and a bust component, which are
characterized by
the shapes of the first (boom) and second (bust) eigenfunctions. The
functional principal components that correspond to the strength of boom and
bust, respectively, are uncorrelated (Figure~\ref{figurewarppc12scores}). Thus, the price swings observed
over the last 15 years can be classified into four distinct categories, markets
that exhibited primarily a bust component (Detroit) and that are contrasted
with markets with primarily a boom (Washington, DC) and, on the other hand,
markets that experienced a combined strong boom and bust cycle (Miami) and
those with relatively weaker booms and busts (Denver). The latter is likely
due to these market's strong underlying growth rates. This means that a bust
phase cannot be clearly predicted from the strength of a preceding boom
in our
quantification, and therefore is hard to predict not only in its timing but
also in terms of whether it will happen or not. There are also a few ``middle
of the road'' markets where the observed time warping is close to the overall
mean time warping, which includes Chicago, San Diego and Minneapolis.

Our analysis also provides some indication that the strengths of booms and
busts may be related to the underlying ``rational'' steady rate of
growth, in
the sense that lower underlying growth rates are associated with larger bust
components and likely also with larger booms (Figure~\ref{figurepcvsalpha}). Such relationships
might reflect the preferences of investors, who primarily may have
invested in
the slower-growing markets, where larger subsequent booms and thus profits
seemed plausible. Such an influx of investors would first drive prices
up and
then down when the investors leave these markets.

A central feature of our approach to analyze asset price dynamics is
the new
concept of nonmonotone warping functions. All existing time-warping approaches
are based on monotonic warping functions, satisfying the constraint
that time always flows forward. A
primary motivation of requiring monotonicity is to preserve the order
of events
(such as landmarks). For our approach the emphasis, however, is not to preserve
the forward flow of time, but rather to allow for reversals of time during
periods of price deflation, implementing the notion that in such
periods prices
retreat to those of an earlier time.

For the existing warping approaches, difficulties arise from the fact
that the
warping function and the amplitude function usually are not jointly
identifiable. Traditionally, this problem has been addressed by
imposing a
variety of identifiability constraints [\citet
{GerGas05,knei08}], none of which
is particularly well motivated. Such constraints are usually imposed for
technical reasons, are hard to verify and may lead to difficulties in
interpreting the results. In the proposed approach, we bypass the
identifiability issue through explicitly exploiting the dependency of the
warping function on the observations. This requires the presence of an
undisturbed interval, on which market price can be reasonably assumed
to follow
a simple smooth trajectory. Both the actual housing data analysis and
simulations show that the choice of such an interval is not critical. The
resulting warping functions are viewed here as a mechanism that gives
rise to
complex features and oscillations in the asset price curves through phase
variation in the underlying dynamics (we assume a simple linear dynamical
system).

Unlike most existing methods, the warping functions in our approach are not
constrained. Particularly, they are not required to be monotonic. When a
warping function is decreasing, we interpret this as a reverse time effect,
that is, the corresponding system is going backward in time, while
periods with
increasing warping functions correspond to phases where the system is going
forward in time. We emphasize that curve alignment is not our goal and,
therefore, warping function monotonicity is not an issue. Rather, the warping
functions themselves are objects of interest since they serve to
quantify the deviations
from smoothly increasing asset prices.

The assumption of an underlying linear dynamical system seems
reasonable for
asset price modeling. This assumption can be fairly easily replaced by another
form of underlying (nonlinear of known shape) dynamical system and the results
can be extended to such more general cases. The system (linear or more complex)
will be fitted on a relatively undisturbed interval, where one may reasonably
assume that price oscillations are absent or are minimal. We need to assume
that such a period of time exists, but a priori knowledge of its exact location
is not required, as the simulations clearly show.

While for the housing index data the simple linear dynamics has great
appeal and provides many insights into the dynamics, it can be
occasionally of interest for other applications such as demography or
biological weight growth to consider an extension of our assumption
that the growth rates $\alpha_i$ do not depend on time or age $t$.
To check whether indeed this may be the case in such applications, one
may consider a second (or higher) order dynamic system.
If, for example, as before, $X(t)=Z(h(t))$ but now
$Z'(t)/Z(t)=\alpha(t)$, then it is easy to see that
\[
\frac{X''(t)}{X(t)}=h'(t)^2\bigl[\alpha'
\bigl(h(t)\bigr) + \alpha^2\bigl(h(t)\bigr)\bigr] +
h''(t)\alpha\bigl(h(t)\bigr),
\]
which can be alternatively expressed as
\[
\frac{d}{dt} \frac{X'(t)}{X(t)}=h''(t) \alpha
\bigl(h(t)\bigr) + h'(t)^2 \alpha '
\bigl(h(t)\bigr).
\]
Such relationships can be used as a diagnostic, by first fitting the
original model and then checking whether indeed
$\frac{d}{dt} \frac{X'(t)}{X(t)}=h''(t) \alpha$ as being implied by
the original model.

Since the rate parameters $\alpha_i$ of the underlying linear dynamic systems
are market-specific and thus differ across the observed curves, they
provide a
complete description of the smooth price increases that are
complemented by the
time warping. Here the baseline growth trajectories $\exp(\alpha_i t)$
themselves correspond to the ``aligned curves,'' which are thus characterized
by one random growth rate. The signal of interest that relates to the
``irrational'' price swings is in the time warping functions and, accordingly,
their functional principal component analysis characterizes the main features
of interest, the modes of variation, which are of special interest for the
recent boom--bust cycle of the housing markets, as we have demonstrated
with the
housing price analysis. Another interesting application is to combine
time warping with clustering of (sub-)metropolitan areas. This may be
achieved through clustering of the PC scores of warping functions as
estimated in this paper.
Moreover, specific features of time-warping functions $h_i$ can be more
generally used for clustering, for example, \citet
{liu2009simultaneous,claeskens2010phase,tang2009time}. Such time
clustering will be of interest for the analysis of housing markets,
either entire metropolitan markets or sub-metropolitan areas within
such markets, depending on data availability.

\begin{appendix}\label{app}
\section*{Appendix: Details on functional principal component~analysis}

In the following the warping functions $h_i,   i=1,\ldots,n$, that are
derived from equation (\ref{eqwarpraw}), are considered to be an
i.i.d. sample of
an underlying time-warping process $H$, defined on the interval
$[0,1]$. For
the functional principal component representation of the process $H$,
the key
components are the mean function $\mu(t)=E(H(t)),   t \in[0,1]$, and the
auto-covariance function $G(s,t)=\operatorname{cov}(H(s), H(t)),  s,t
\in[0,1]$.

We assume throughout that the time-warping process $H$ has smooth and square
integrable trajectories, $E\int_{[0,1]} H^2(t)\,dt < \infty$. Then the
eigenfunctions $\phi_k,   k \ge1$,\vadjust{\goodbreak} of the auto-covariance operator
$A(f)(t)=\int_{[0,1]} f(s)G(s,t)\,ds,  t \in[0,1]$, which is a linear
Hilbert--Schmidt operator that maps $L^2([0,1])$ into itself, form the
orthonormal
eigenbasis. Processes $H$ may then be represented in this basis by
means of the
Karhunen--Lo\`eve expansion [\citet{ash75}]
\[
H(t)=\mu(t) + \sum_{k \ge1} \xi_k \sqrt{
\lambda}_k\phi_k(t),\qquad t \in[0,1],
\]
where the $\xi_k\sqrt{\lambda}_k$ are the functional principal components,
which correspond to projections $ \xi_k \sqrt{\lambda}_k=\int_{[0,1]} (H(t)-\mu(t))\phi_k(t)\,dt$,
and satisfy $E(\xi_k)=0$ and $E(\xi_k^2)=1$, with $\lambda_k$
denoting the $k$th
eigenvalue. In the following, the $L^2$ norm of a function $f$ will be denoted
by
$\llVert f\rrVert _{L^2}=[\int_{[0,1]} f^2(t) \,dt]^{1/2}$ or $\llVert
f\rrVert _{L^2}=[\int_{[0,1]} \int_{[0,1]} f^2(s,t) \,ds\,dt]^{1/2}$.\vspace*{2pt}

The empirical estimates of the mean and auto-covariance functions of
$H$ are
%
\begin{eqnarray}
\hat{\mu}(t) &=& \frac{1}{n}\sum_{i=1}^{n}
h_i(t ),\qquad t\in[0,1], \label{EmpMu}
\\
\hat G(s, t) &=& \frac{1}{n}\sum_{i=1}^{n}
h_i(s )h_i(t ) - \hat\mu(s )\hat\mu(t ),\qquad s,t
\in[0,1]. \label{EmpCov}
\end{eqnarray}
From\vspace*{1pt} the fact that $E(\hat{\mu})=\mu$ and the square integrability, one
immediately finds $E\llVert \hat{\mu}-\mu\rrVert _{L^2}^2=\frac
{1}{n}\int_{[0,1]}\operatorname{var}(H(t))\,dt$
and, therefore, $\llVert \hat{\mu}-\mu\rrVert
_{L^2}=O_p(n^{-1/2})$. Using more intricate
arguments, it can been shown that under the additional assumption that
$E\llVert H^2\rrVert _{L^2}^2<\infty$, it also holds that $\llVert
\hat{G}-G\rrVert _{L^2}=O_p(n^{-1/2})$
[\citet{daux82,hall061}]. Results of the above type for $\hat
{\mu}$ and
$\hat{G}$ are then typically coupled with a perturbation result that relates
differences in eigenfunctions and eigenvalues to those of $L^2$
distances of
covariances; an example is Lemma~4.3 in \citet{bosq00}, which is
utilized to
prove the following more rigorous results about the convergence in sup
norm of
the eigenfunctions and eigenvalues of the time-warping process $H$.

Such results can be obtained by adopting an approach of \citet
{mull123} that
utilizes techniques developed in \citet{li10}. Technical
assumptions needed are
as follows: For arbitrary universal constants $0<B<C<\infty$, it holds
that for
all $t$ that $E|H(t)|^\ell\leq\frac{\ell!}{2} C^{\ell-2}B^2, \ell
= 2,3,\dots$ and for all $s,t$ that $E|H(s)H(t)|^\ell\leq\frac{\ell!}{2}
C^{\ell-2}B^2, \ell= 2,3, \dots.$ In addition, all trajectories of
$H$ are
assumed to satisfy $|H(s)-H(t)| \leq C|s-t|$\break and
$E(\sup_{t\in\mathcal{T}}|H(t)|) < \infty$,
$\sup_{t\in[0,1]}|\mu(t)| < \infty$, $\sup_{s,t}|G(s, t)| <\infty
$,\break $\sup_{t\in[0,1]}|\phi_k(t)| < \infty$ for each\vspace*{1pt} $k\geq1$.

Under these assumptions, using similar arguments, such as those
provided in Lemma~1 and Theorem~1 of \citet{mull123},
leads to the following\vadjust{\goodbreak} consistency results:
\begin{eqnarray*}
\sup_{t \in\mathcal{T}}\bigl|\hat{\mu}(t )- \mu(t)\bigr|& =& O\bigl((\log n/
n)^{1/2}\bigr)\qquad\mbox{a.s.},
\\
\sup_{s, t \in\mathcal{T}}\bigl|\hat{G}(s, t) -G(s, t)\bigr| &=& O\bigl((\log n/
n)^{1/2}\bigr)\qquad\mbox{a.s.},
\\
\sup_{t \in\mathcal{T}}\bigl|\hat{\phi_k}(t)- \phi_k(t)\bigr|
&=& O\bigl((\log n/ n)^{1/2}\bigr)\qquad\mbox{a.s.},
\\
|\hat{\lambda}_k - \lambda_k| &=& O\bigl((\log n/
n)^{1/2}\bigr)\qquad\mbox{a.s.}
\end{eqnarray*}
Here $\hat{\phi}_k$ and
$\hat{\lambda}_k$ are the eigenfunction/eigenvalue estimates that one obtains
from a spectral decomposition of the empirical covariance function
$\hat{G}$ in
(\ref{EmpCov}). This is numerically implemented through suitable discretization
and using matrix eigenanalysis. We note that the consistency of the estimates
$\xi_{ik} \sqrt{\lambda}_k=\int_{[0,1]} (h_i(t)-\hat{\mu}(t))\hat
{\phi}_k(t))\,dt$ of the
$k$th principal components $\xi_{ik} \sqrt{\lambda}_k,  i=1,\ldots,n,  k \ge1$, follows
immediately from these results.
\end{appendix}

\section*{Acknowledgments}
We thank anonymous reviewers and editors for
comments that led to significant improvements of this paper.

\begin{supplement}
\sname{Supplementary material}
\stitle{Time-warped growth processes with applications to the modeling of boom--bust cycles: Additional analysis\\}
\slink[doi]{10.1214/14-AOAS740SUPP} 
\sdatatype{.pdf}
\sfilename{AOAS740\_supp.pdf}
\sdescription{We provide some additional analysis of the housing index data.}
\end{supplement}



\printaddresses
\end{document}